\documentclass[a4paper]{article}
\usepackage{epsfig,amsmath,amsfonts,amssymb,setspace,multirow,textcomp}
\usepackage[T1]{fontenc}
\makeatletter
\renewcommand{\@evenfoot}{\hfil \thepage \hfil}
\renewcommand{\@oddfoot}{\hfil \thepage \hfil}
\makeatother

\renewenvironment{thebibliography}[1]{\begin{oldthebibliography}{#1}\setlength{\parskip}{0ex}\setlength{\itemsep}{0ex}}{\end{oldthebibliography}}
\begin{document}
\fontsize{11}{11}\selectfont % the font size cannot be changed in any case!
%  insert your title, authors information and text instead of the one provided below
%\title{HIGH-SPEED MULTICOLOR PHOTOMETRY WITH CMOS CAMERAS}

%%%%%%%%%%%%%%%%%%%%%%%%%%%%%%%%%%%%%%%%%%%%%%%%%%%%%%%%%
%%%\noindent {\small{\it Submitted to the Proceedings of the Conference "22nd Annual Student
%%%Conference, Week of Doctoral Students 2013", Charles University in Prague, June 4–7, 2013}}

%%%\hrule
%\medskip \hrule \medskip

%\bigskip
%\bigskip
%%%%%%%%%%%%%%%%%%%%%%%%%%%%%%%%%%%%%%%%%%%%%%%%%%%%%%%%

\title{Performance of commercial CMOS cameras for high-speed \\ multicolor photometry}

\author{\textsl{S.\,M.~Pokhvala$^{1}$, B.\,E.~Zhilyaev$^{1}$, V.\,M.~Reshetnyk$^{2}$}}
\date{\vspace*{-6ex}}
\maketitle

\begin{center} {\small $^{1}$Main Astronomical Observatory, NAS of Ukraine, Zabalotnoho 27, 03680, Kyiv, Ukraine\\
$^{2}$Taras Shevchenko National University of Kyiv, Glushkova ave., 4, 03127, Kyiv, Ukraine\\
{\tt nightspirit10@gmail.com}}
\end{center}

\begin{abstract}
We present some results of testing of commercial color CMOS cameras for astronomical
applications. CMOS sensors allow to perform photometry in three filters simultaneously that
gives a great advantage compared with monochrome CCD detectors. The Bayer BGR colour system
realized in CMOS sensors is close to the Johnson BVR system. We demonstrate transformation
from the Bayer color system to the Johnson one.  Our photometric measurements with color
CMOS cameras coupled to small telescopes (11 - 30 inch) reveal that in video mode stars up
to V $\sim$ 9 can be shot at 24 frames per second. Using a high-speed CMOS camera with
short exposure times (10 - 20 ms) we can perform an  imaging mode called "lucky imaging".
We can pick out  high quality frames and combine them into a single image using
"shift-and-add" technique. This allows us obtain an image with much higher resolution than
would be possible shooting a single image with long exposure.  For image selection we use
the Strehl-selection method.  We demonstrates advantage of the lucky imaging technique in
comparison with long exposure shooting. The FWHM of the blurred image caused by atmosphere
turbulence can be decreased twice and more.


\end{abstract}

%\section{Introduction}
\section*{\sc introduction}

\indent \indent  Multicolor synchronous observations in astronomy are important for the
study of transient events.  Traditional methods lose a signal due to the serial
measurements in different filters. Multicolor sensors based on metal-oxide semiconductors
(CMOS) allow refusing the use of filters in general \cite{butler08}.

Commercial CMOS cameras provide simultaneous imaging of an object in the Bayer color
system: the blue filter "B", ($\lambda > 400$ nm), the red "R" ($\lambda < 900$ nm) and the
intermediate filter "G". Transformation of the Bayer color system BGR to the international
Johnson BVR system is a relatively easy problem, as can be shown below. The inconvenience
of CMOS sensors is comparatively low quantum efficiency. However, it plays a role only for
extremely faint stars. Commercial CMOS cameras allow us to perform fast high-precision
multicolor photometry of relatively bright objects with small telescopes. As an example we
can refer to photometric records of stars in Trapeze of Orion with the 60 cm telescope in
the white light at Andrushivka observatory in 2003.  The commercial video camera SANIO,
VCB-3572IRP was employed.  Objects: MT Ori, V1229 Ori, V1333, Ori AD Ori (V $\sim$ 11.2,
11.4, 12.2, 12.9). The exposure time was 0.02 s, frequency of 15 fps.  AD Ori (V = 12.9)
demonstrates photometric precision of $\sim$ 0.1 magnitude.

Using high values of ISO, 6400 and above provides an opportunity to capture objects up to V
$\sim$ 10 with a frequency up to 24 fps with small telescopes. This allows us to study
high-frequency variability in the range of 10 Hz and above in three colors simultaneously.

As shown by authors in \cite{Pokhvala} exposures of 30 seconds are capable of reaching
magnitudes 16.5 with a S/N = 3, using 11 inches of aperture with a commercial camera Nikon
D90.  The S/N = 3 represents a lower limit of detection of an object with error of 0.3
magnitude.  Again 19.5 magnitude objects are detectable with a sky of better 20th magnitude
per square arc-second, a S/N = 3, and an exposure of one hour, using 11 inches of aperture
only.

Thus, the testing results reveal that commercial CMOS cameras can create serious
competition with modern CCD cameras in astronomy.

\section*{\sc Transformation coefficients}

\indent \indent  The Bayer BGR colour system realized in CMOS sensors is close to the
Johnson BVR system. To verify these transformation coefficients were determined from
comparison star observations using the following equation proposed in \cite{Harris}
\begin{equation}\label{}
    v=V+a_{1}+a_{2}X+a_{3}(B-V)+a_{4}(B-V)^{2}
\end{equation}
where $v$ is the instrumental magnitude in the green channel of the Bayer color system, B,
V are the standard Johnson magnitudes and X is the air mass, $a_{1} - a_{4}$ are the
transformation coefficients. Then we perform a Least Squares solution for the coefficients.

Observations were carried out using a commercial Canon 350D camera on the 0.7 m telescope
from urban site at Main Astronomical Observatory in Kiev on two nights 2012 October 12-13.

The standard deviations of fit for 5 stars as can see in Fig. 1 were 0.05 in V. Field stars
around V390 Aur were used as standard stars. The comparative high standard deviations
result from using not standard stars from the TYCHO survey with magnitudes known to
0.03-0.1. These measurements simply provide the information about the precision of the
measurement. However, we can conclude that our transformed magnitudes are within errors of
the TYCHO survey.

So transformation of the Bayer color system BGR to the international Johnson BVR system can
be easy performed.

%%%%%%%%%%%%%%%%%%%%%%%%% Fig 1, 2
\begin{figure}[!h]
\centering
\begin{minipage}[t]{.45\linewidth}
\centering
\epsfig{file = 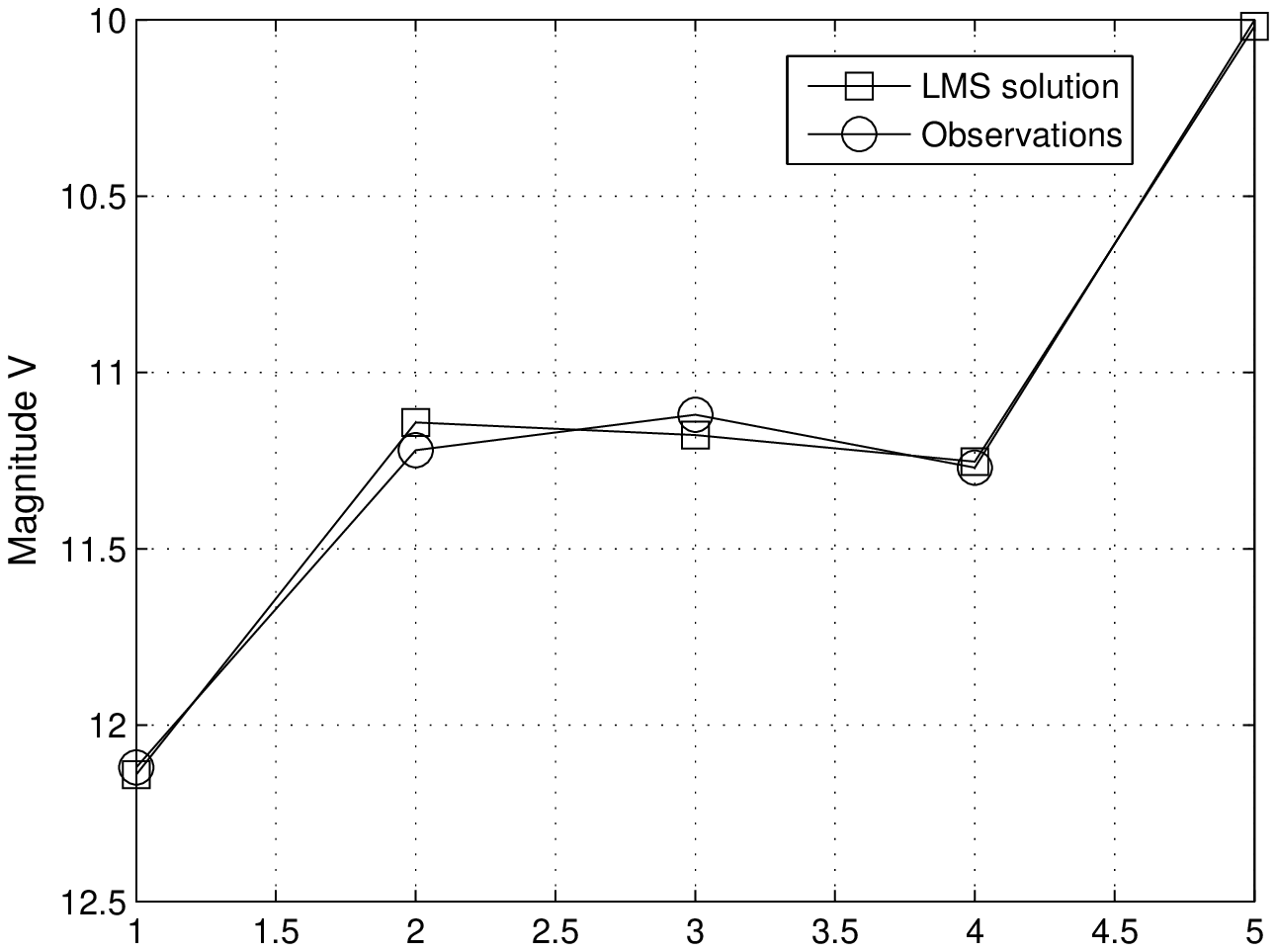,width = .8\linewidth} \caption{The Bayer G colour (squares)
transformed to the Johnson V color (circles).}\label{fig1}
\end{minipage}
\hfill
\begin{minipage}[t]{.45\linewidth}
\centering
\epsfig{file = 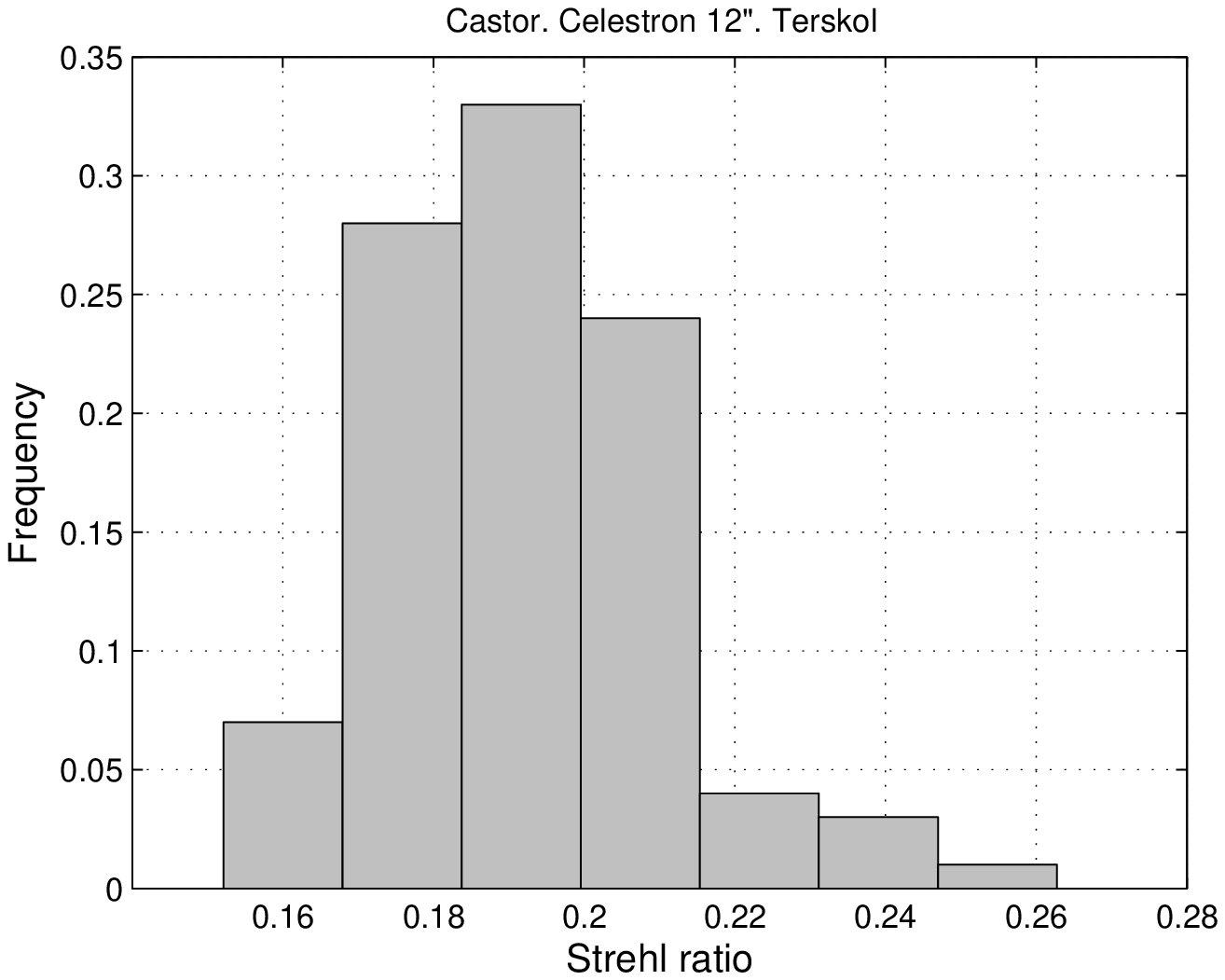,width = .8\linewidth} \caption{Probability density of Strehl
ratio of Castor.}\label{fig2}
\end{minipage}
\end{figure}

%%%%%%%%%%%%%%%%%%%%%%%%%

%%%%%%%%%%%%%%%%%%%%%%%% Fig 3, 4
\begin{figure}[!h]
\begin{minipage}[t]{.45\linewidth}
\centering
\epsfig{file = 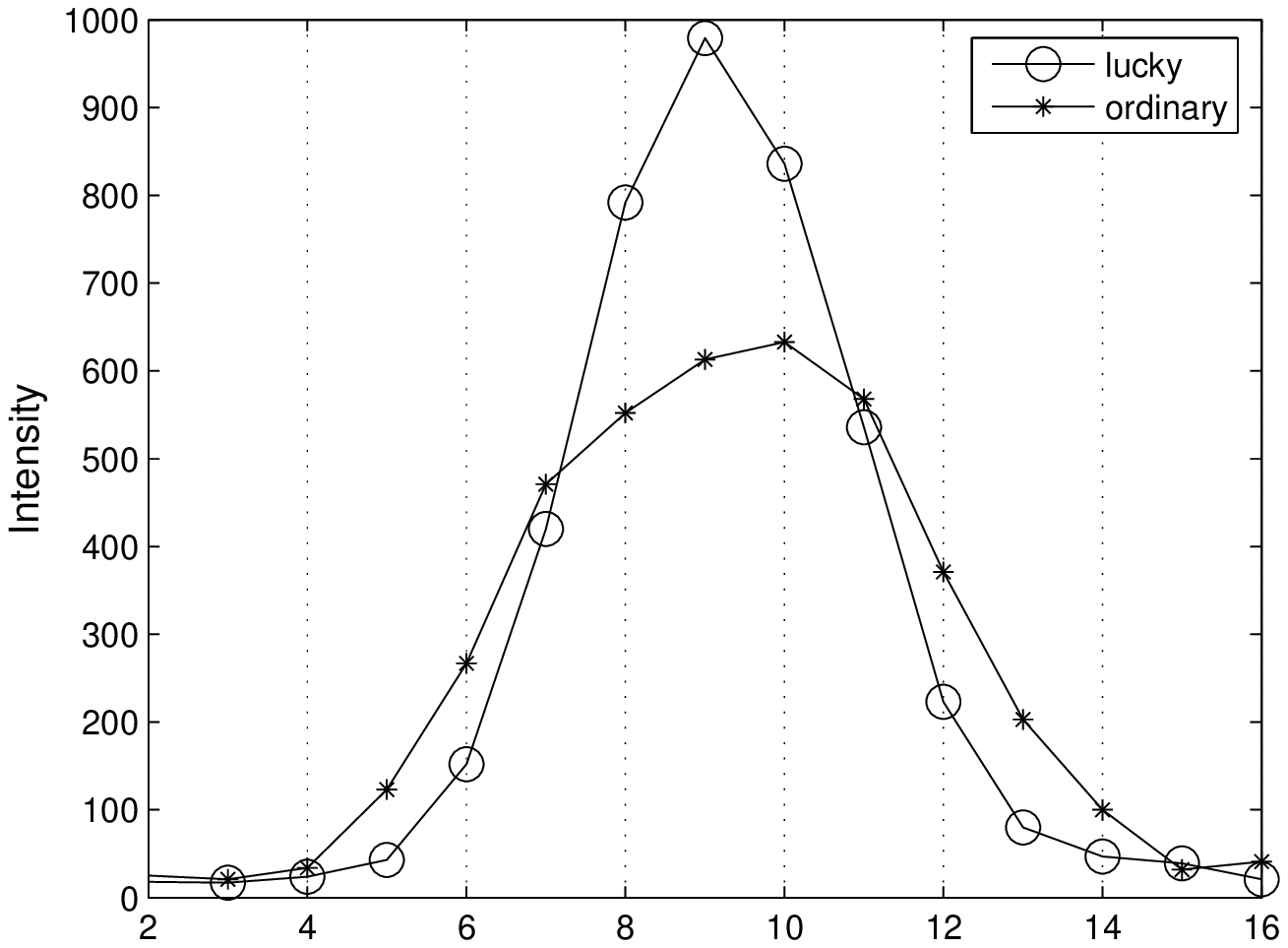,width = .8\linewidth} \caption{The photometric profiles  of
the blurred image (FWHM = 3.46 arcsec) and of the lucky one (FWHM = 2.50
arcsec).}\label{fig3}
\end{minipage}
\hfill
\begin{minipage}[t]{.45\linewidth}
\centering
\epsfig{file = 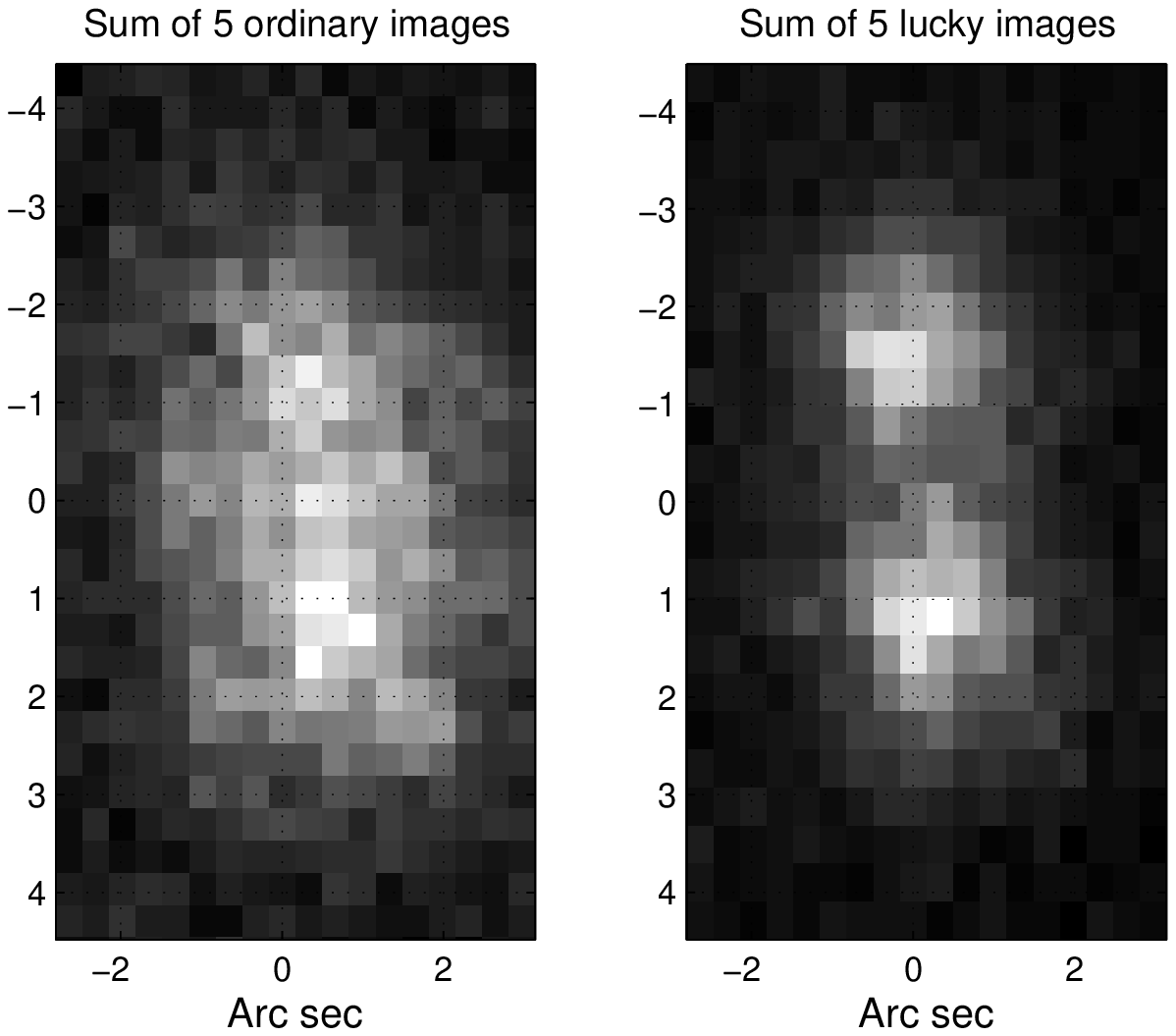,width = .8\linewidth} \caption{Shows $eps^{1}$ Lyr imaged
with the 70 cm telescope using the lucky imaging technique.}\label{fig4}
\end{minipage}
\end{figure}

%%%%%%%%%%%%%%%%%%%%%%%%%%

\section*{\sc High-frequency observations with CMOS cameras}

\indent \indent As shown by authors in \cite{Pokhvala} the continuous shooting
with a frequency 24 fps with 11 inches of aperture equipped with a commercial camera Nikon
D90 allows to study rapid variability of stars up to 7 mag with high photometric precision
and of 9 mag with acceptable accuracy. This opens the way to study the rapid variability of
the large number of stars, among which there are many objects, which gave the name to the
prototypes of stellar variability.

Using a high-speed CMOS camera with short exposure times (10 - 20 ms) we can perform a form
of speckle imaging called "lucky imaging". We can pick out  high quality frames ("lucky
images") and combine them into a single image using "shift-and-add" technique. This allows
us obtain an image with much higher resolution than would be possible with long exposure
shooting.

From a series of images taken at a speed of 24 fps, five images were chosen with high
Strehl ratio for test purpose. Additionally, five images with low Strehl ratio were
stacked. Fig. 2 presents histogram of Strehl ratio of Castor imaged with the Celestron 11"
telescope equipped with a Nikon D90 camera at Peak Terskol. Fig. 3 demonstrates advantage
of the lucky imaging technique in comparison with long exposure shooting. The advantage is
evident from this illustration. The FWHM (the full width at half maximum) of the blurred
image is estimated to be equal 3.46 arcsec.  At the same time the FWHM of the selected
images is around 2.50 arcsec.

For image selection we use the Strehl-selection method \cite{Sacek}. By
definition, Strehl ratio  is the ratio of peak diffraction intensities of an aberrated
versus perfect wavefront. The ratio indicates the level of image quality in the presence of
wavefront aberrations caused by atmosphere turbulence. For example, 0.20 Strehl indicates
about 80\% lower energy within the Airy disc.

Fig. 4 demonstrates quite remarkable result: a double system can be clearly detected with a
small aperture under bad atmospheric condition (seeing 3.97 arcsec). The separation between
the two stars is around 2.35 arcsec and the FWHM of the individual stars is around 1.70
arcsec. One needs a relatively bright star in the field of view on which Strehl ratio can
be evaluated. As shown in \cite{Pokhvala} the continuous shooting
with 11 inches of aperture equipped with a commercial camera allows to operate with a guide
star up to 9 mag with acceptable accuracy

\section*{\sc Conclusion}

\indent \indent  Testing of commercial color CMOS cameras for astronomical applications demonstrates that
they allow to carry out photometry in three filters simultaneously that gives an obvious
advantage compared with monochrome CCD detectors.

We demonstrate that the Bayer BGR colour system realized in CMOS sensors can be easily
transformed to the Johnson BVR system. Photometric measurements with color CMOS cameras
coupled to small telescopes (11 - 30 inch) reveal that in video mode stars up to V $\sim$ 9
can be shot at 24 frames per second.

We performed a "lucky imaging" mode using the Strehl-selection method and "shift-and-add"
technique to combine high quality frames into a single image.  This allows us obtain an
image with much higher resolution than would be possible shooting a single image with long
exposure.  We demonstrates advantage of the lucky imaging technique in comparison with long
exposure shooting. The FWHM of the blurred image caused by atmosphere turbulence can be
decreased twice and more.

\section*{\sc acknowledgement}
\indent \indent
The authors would like to particularly thank Maxim Andreev at the Peak Terskol Observatory for help during the observations.

%

%\begin{references}

%\footnotesize
%\reference Klimov, S. {\it et al.}, ASPI experiment: Measurements of fields and waves
%onboard the Interball-1 spacecraft, {\it Ann. Geophys.}, 15, 514-527, 1997.

\end{document}